\documentclass[manuscript]{aastex}

\usepackage{epsfig}
\usepackage{bm}
\usepackage{subfigure}
\usepackage{multicol}
\usepackage{multirow}
\usepackage{array}
\newcommand{\PreserveBackslash}[1]{\let\temp=\\#1\let\\=\temp}
\newcolumntype{C}[1]{>{\PreserveBackslash\centering}p{#1}}
\newcolumntype{R}[1]{>{\PreserveBackslash\raggedleft}p{#1}}
\newcolumntype{L}[1]{>{\PreserveBackslash\raggedright}p{#1}}

\shortauthors{Song \& Zhang} \shorttitle{Effects of non-radial field on measuring helicity transport}

\begin{document}

\title{Effects of non-radial magnetic field on measuring magnetic helicity transport across solar photosphere}

\author{Y. L. Song\altaffilmark{1,2} \& M. Zhang\altaffilmark{1}}

\altaffiltext{1}{Key Laboratory of Solar Activity, National Astronomical Observatories, Chinese Academy of
 Sciences, A20 Datun Road, Chaoyang District, Beijing 100012, China; \\ Email: ylsong@bao.ac.cn}
\altaffiltext{2}{University of Chinese Academy of Sciences, China}

\begin{abstract}
It is generally believed that the evolution of magnetic helicity has a close relationship with solar activity. Before the launch of SDO, earlier studies have mostly used MDI/SOHO line of sight magnetograms and assumed that magnetic fields are radial when calculating magnetic helicity injection rate from photospheric magnetograms. However, this assumption is not necessarily true. Here we use the vector magnetograms and line of sight magnetograms, both taken by HMI/SDO, to estimate the effects of non-radial magnetic field on measuring magnetic helicity injection rate. We find that: {\it 1}) The effect of non-radial magnetic field on estimating tangential velocity is relatively small; {\it 2}) On estimating magnetic helicity injection rate, the effect of non-radial magnetic field is strong when active regions are observed near the limb and is relatively small when active regions are close to disk center; {\it 3}) The effect of non-radial magnetic field becomes minor if the amount of accumulated magnetic helicity is the only concern.
\end{abstract}



\section{Introduction}

Magnetic helicity is a physical quantity that describes the topology and complexity of a magnetic field. It is believed that magnetic helicity plays an important role in solar activity (Zhang \& Low 2005). Theoretical studies have suggested that the accumulation of magnetic helicity in the solar atmosphere provides free magnetic energy; since the total magnetic helicity in a force-free magnetic field has an upper bound, once this limit is exceeded, a non-equilibrium situation will then result in CME expulsion (e.g. Zhang et al. 2006, Zhang \& Flyer 2008, Zhang et al. 2012).

Observations also suggest that impulsive changes of the magnetic helicity injection rates have occurred during solar flares (Moon et al. 2002a, 2002b, 2003; Zhang et al. 2008). In a survey on 393 active regions, Labonte et al. (2007) suggested that X-class flares occurred when the peak magnetic helicity injection rate is greater than $6\times10^{36}{Mx}^2s^{-1}$. Nindos and Andrews (2004) suggested that it is the amount of stored magnetic helicity that will determine whether a big flare will be eruptive or confined. Park et al. (2010) studied the occurrence of the X3.4 flare in NOAA 10930 and concluded that the flare may be initiated by a helicity injection, into a system of oppositely signed helicity. Similarly, Vemareddy et al. (2012) studied the helicity injection in two active regions and suggested that flux motions and the spatial distribution of helicity injection are important ingredients in the understanding of the favorable conditions for solar eruptions. All these indicate that magnetic helicity plays an important role in solar activities and hence it is important to have an accurate measurement of the magnetic helicity injection rate.

The rate of magnetic helicity flux across solar surface \textsl{S}, that is, the helicity injection rate, can be estimated by the following equation (Berger 1984):
$$ \frac{dH_R}{dt}\bigg|_s =2\int_{s}[(\mathbf{A}_p\cdot \mathbf{B}_t)v_n-(\mathbf{A}_p\cdot \mathbf{V}_t)B_n]ds,\eqno{(1)}$$
\noindent where $\mathbf{B}_t$ and $B_n$ are the tangential and normal magnetic fields, $\mathbf{V}_t$ and $V_n$ are the corresponding velocities. $\mathbf{A}_p$ is a unique vector potential satisfying following conditions:
$$\nabla\times\mathbf{A}_p\cdot\hat{\textbf{n}}=B_n,~~\nabla\cdot\mathbf{A}_p =0,~~\mathbf{A}_p\cdot\hat{\textbf{n}}=0.\eqno{(2)}$$

To apply Equation (1) to observational data, we need to know $\mathbf{B}_t$, $B_n$, $\mathbf{V}_t$ and $V_n$. However, before the launch of SDO, the only available data with good temporal-resolution and continuous, uniform observations are longitudinal MDI/SOHO magnetograms. To make use of these data to estimate helicity injection rate $ \frac{dH_R}{dt}$, two hypothesis are frequently invoked.

The first one was suggested by D\'{e}moulin \& Berger (2003). They argued that the horizontal motions ($\textbf{U}$), deduced by tracking the footpoints of magnetic flux tubes, already include the effect of both the emergence and the shearing motions. This is to say that, the horizontal velocity \textbf{U} obtained by tracking the movement of magnetic footpoints is actually
$$\textbf{U}=\textbf{V}_t - V_n\cdot\frac{\textbf{B}_t}{B_n} ~~.\eqno{(3)}$$
\noindent With this, Equation (1) becomes simplified as
$$\frac{dH_R}{dt}\bigg|_s = -2\int_{s}(\mathbf{A}_p\cdot\mathbf{U})\cdot B_n ds ~~.\eqno(4)$$
\noindent This approach has since then become a standard procedure in calculating helicity injection rate, although \textbf{U} can be obtained by various methods such as LCT (e.g. Chae 2001), DAVE (e.g. Schuck 2006) and NAVE (e.g. Schuck 2005) etc. Recent work (Schuck 2008; Liu and Schuck 2012; Kazachenko et al. 2014; Liu et al 2014) has called into question the validity of the Demoulin and Berger conjecture in which the apparent horizontal motions determined by e.g. LCT flows include both the twisting and flux emergence terms. These studies find instead that LCT or DAVE-derived flows correspond mainly with the true horizontal motions, and are insensitive to flux emergence. On the other hand, Liu and Schuck (2012) and Liu et al (2014) find that in a variety of active regions, the horizontal (twisting) term dominates the flux emergence term, even in cases where magnetic flux is observed to be emerging.  The end result is that a straightforward application of Equation (4), in which the velocity \textbf{U} used include the results of horizontal velocities returned by LCT or DAVE, should include the dominant contribution to the magnetic helicity flux.

The second hypothesis comes from assuming that magnetic fields on the photosphere are predominantly vertical. With this, the vertical magnetic field strength $B_n$ in Equations (1) - (4) can then be estimated as the observed line of sight magnetic field strength $B_l$ times $ 1 / \cos\theta $, where $\theta$ is the heliocentric angle of the region. Using active region NOAA 10365 as an example, Chae et al. (2004) argued that this method can be more generally applied, even to regions with inclined magnetic fields. Similar to the approach proposed by D\'{e}moulin \& Berger (2003), this has also become a standard approach and many studies have been carried out based on these two hypothesis.

With the launch of SDO, we now have the full disk vector magnetograms with good time cadence and continuity (Hoeksema et al. 2014, Bobra et al. 2014). Now we can use Equation (1) to calculate the full terms of helicity injection rate. At the same time, it is possibly also the time for us to find out the degree of inaccuracy that our above two hypothesis have brought to us, in order to better understand previous results that have been obtained based on the two hypothesis.

In this paper we intend to investigate the influence of the second hypothesis, that is, the effect of the existence of non-radial magnetic field on measuring magnetic helicity transport. We organize our paper as follows. In Section 2, we present the sample and data reduction. Our analysis and results are given in Section 3. Conclusion and discussion are presented in Section 4.


\section{Sample and data reduction}

The magnetograms we used in this study are taken by HMI/SDO. The HMI instrument (Scherrer et al. 2012; Schou et al. 2012a, 2012b) observes the full solar disk at 6173 \r{A} with a $4096 \times 4096$ CCD detector to study the oscillations and magnetic fields on the photosphere. The magnetograms are obtained using a Milne-Eddington (ME) based inversion code (Borrero et al. 2011) and the $180^\circ$ ambiguity problem is resolved using a ``minimum energy" algorithm (Metcalf 1994; Metcalf et al. 2006; Leka et al. 2009).

Two streams of active-region vector data have been produced and provided (Sun et al. 2013), namely, hmi.sharp\_720s and hmi.sharp\_cea\_720s. The series of hmi.sharp\_720s preserves vector data in the native coordinate, i.e. a 2D array as measured at each CCD pixel. The field vectors are expressed as field strength ($B$), inclination ($\gamma$) and azimuth ($\varphi$). The series of hmi.sharp\_cea\_720s uses a Cylindrical Equal Area (CEA) projection. The standard CEA coordinates (x,y) relate to the Heliographic longitude and latitude, with each pixel representing a same area, which makes them suitable for the computation of total flux. The field vectors in hmi.sharp\_cea\_720s are presented as ($B_r$, $B_\theta$, $B_\phi$) in heliocentric spherical coordinate, and are obtained by mathematical methods, called ``remapping'' and ``vector transformation'', from the hmi.sharp\_720s data.

We use both streams of the active-region vector data in this study. We use series of hmi.sharp\_cea\_720s to obtain $B_r$, that is, $B_n$ in Equations (1) - (4). We use series of hmi.sharp\_720s to get the longitudinal magnetic field $B_l$, the type of data used in most previous studies. To make the two series of data comparable on a pixel-to-pixel basis, we have remapped the hmi.sharp\_720s data to CEA coordinate.

Since we are primarily interested in testing the second hypothesis, we will focus on comparing the difference between using the ``real'' $B_n$, from vector data, and the ``estimated'' $B^{\prime}_n$, by assuming that the magnetic field is radial. We will use Equation (4) to calculate the magnetic helicity injection rate. The vector potential $\mathbf{A}_p$ and the horizontal velocity field \textbf{U} are derived by the methods of fast Fourier transform (FFT) and the Differential Affine Velocity Estimator (DAVE; Schuck, 2006) respectively. The accumulated magnetic helicity is also calculated as
$$H(t)=\int^{t}_{0}\frac{dH_{R}(t)}{dt}dt ~~~,\eqno{(5)}$$
\noindent where $t=0$ is the starting time of our calculation.

We studied three active regions, NOAA 11072, NOAA 11084 and NOAA 11158. Both NOAA 11072 and NOAA 11158 are emerging active regions and have been studied by Liu \& Schuck (2012) with a purpose different from ours.  NOAA 11072 has a bipolar magnetic field structure with no C-class or above flares occurred during its passage of the disk. NOAA 11158 is a multipolar complex active region that produced several major flares. For a comparison, NOAA 11084 is selected as a mature and simple active region with no significant flaring activity. Compared to Liu \& Schuck (2012), our sample contains data extending closer to the solar limb, ending up when the three active regions are 60 degrees in longitude near the western limb.  Table.1 lists some properties of these three active regions.

\begin{table}[htbp]
\centering
\centerline{\footnotesize \textbf{Table 1.} Sample of active regions}
\label{tab1}
{\scriptsize
\begin{tabular}{ccccccccc}
\hline
\hline
NOAA & \multicolumn{2}{c}{Start of Observation} & & \multicolumn{2}{c}{End of Observation} & Sunspot \\
\cline{2-3} \cline{5-6}
No. & Time (UT) & Position & & Time (UT) & Position & Classification \\
\hline
11072 & 2010.05.20\_16:24:00 & S14 E34 & & 2010.05.27\_14:48:00 & S14 W60 & $\beta$  \\
11084 & 2010.06.28\_00:24:00 & S22 E60 & & 2010.07.06\_23:48:00 & S23 W60 & $\alpha$   \\
11158 & 2011.02.10\_22:00:00 & S14 E42 & & 2011.02.18\_14:12:00 & S14 W60 & $\beta\gamma\delta$ \\
\hline
\end{tabular}
}
\end{table}


\section{Analysis and results}

Magnetic field in active regions is not strictly radial, especially in regions such as sunspot penumbra and around magnetic neutral lines. Figure 1 presents the histogram of the ratio between the tangential field ($B_t$) and the normal field ($B_n$) in a magnetogram of NOAA 11072. The magnetogram is obtained at 16:36 UT on 2010 May 25 when the active region is located at S14W33 and only data points whose strengths of normal magnetic field are stronger than 100 G are used. We see that in a significant fraction ($\sim 30\%$) of the active region the strength of the tangential field is larger than that of the radial field. This shows that the tangential field is not ignorable. Note that this is a general situation for most active regions, and the study of how these tangential magnetic fields affect helicity transport is important.

\subsection{On obtaining the radial magnetic field}

As mentioned before, when using MDI/SOHO magnetograms to calculate helicity injection rate, a common practice is to assume that the magnetic field is radial and then obtain the strength of radial magnetic field $B_r$ from the longitudinal magnetic field $B_l$, that is, $B_r^{\prime} = B_l / cos \theta$, where $\theta$ is the heliocentric angle of the observed region. However, this ``estimated'' $B_r$, denoted as $B_r^{\prime}$ hereafter, could be significantly different from the ``real'' $B_r$.

Figures 2, 3 and 4 show a comparison of the estimated $B_r^{\prime}$ and the real $B_r$ for active regions NOAA 11158, 11072 and 11084, respectively. Here we see that when the active region is observed near the disk center, $B_r^{\prime}$ and $B_r$ look quite similar to each other. This indicates that the influence of the tangential field on estimating radial magnetic field is not severe. However, when the active region is observed far away from the disk center, the estimated radial field $B_r^{\prime}$ could be significantly different from the real $B_r$. In some areas (outlined by the red squares in Figures 2 and 3 for example) even the sign of the estimated radial magnetic field $B_r^{\prime}$ could be different from that of the real $B_r$. This will certainly influence the calculation of helicity transport.

To show this difference more quantitatively, we present the scatter plots of $B_r^{\prime}$ vs. $B_r$ in Figure 5 for the three active regions studied. Left panels are for active regions near the disk center and right panels for regions near the limb. The correlation coefficients between $B_r^{\prime}$ and $B_r$ are also presented in the left-top corner of each panel. We see here, when the active regions are near the disk center, the correlation coefficients between $B_r^{\prime}$ and $B_r$ are all higher than 0.95, which means that $B_r^{\prime}$ maps are very similar to $B_r$ maps. When the active regions are at 45 degrees in the west, the correlation coefficients all drop below 0.85. This means that the $B_r^{\prime}$ maps have already be different from $B_r$ maps.

\subsection{Effect on obtaining the tangential velocity field}

Figure 6 shows the tangential velocity field estimated using DAVE method (Schuck 2006) for NOAA 11158.  On using DAVE code, we have followed Liu \& Schuck (2012) and used 19 pixels ($8.5\arcsec$ for HMI data) for window size. Right panels in Figure 6 show the tangential velocities obtained by using $B_r$ magnetograms, and left panels for using $B_r^{\prime}$ magnetograms. Similar to Figure 2, we also show the calculation in two representative moments, one when the active region is near the disk center (top panels) and one when the active region is $45^o$ from the disk center (bottom panels).

It is possibly not surprising to see, from the top panel, that using $B_r^{\prime}$ and $B_r$ give relatively similar maps of tangential velocities, as we have already known from Figure 2 that the influence of the tangential field on estimating radial magnetic field is not severe when the active region is near the disk center. However, what is interesting to find out, from the bottom panels, is that the tangential velocity maps also look similar even though the magnetograms, $B_r^{\prime}$ or $B_r$, based on which the velocities are obtained, are quite different. This means that the effect of non-radial magnetic field on estimating tangential velocity is small, no matter the active region is near or away from the disk center.

To show this more quantitatively, Figure 7 presents the correlation between the two types of velocity maps, one obtained from $B_r$ magnetograms and the other from $B_r^{\prime}$ magnetograms. Top panels are for the component $V_x$, positive in the direction of solar rotation. Middle panels are for the component $V_y$, positive from the south to the north. The bottom panels show the correlation between the magnitudes of the tangential velocities. We see here that in all six panels the correlation is high. The correlation coefficients are all larger than 0.97 when the active region is near the disk center. When the active region is $45^o$ in the west, the correlation coefficients are all above 0.72, still high considering the big differences of the magnetograms based on which they are obtained from.

Similarly, the correlation coefficients between the two types of tangential velocities are calculated and presented in Figure 8, for the three active regions during the entire passage across the disk. We see here that the correlation coefficients are all larger than 0.9 when the active regions are near the disk center. The values of the correlation coefficients decrease when the active regions are moving away from the disk center, which shows that the influence of the non-radial magnetic field is increasing. However, still the correlation coefficients are large enough, compared to those between the helicity injection rates which we will show in the next subsection.

\bigskip
\begin{table}[htbp]
\centering
\centerline{\footnotesize \textbf{Table 2.} Statistics on the correlation coefficients between tangential velocity fields}
\label{tab1}
{\scriptsize
\begin{tabular}{cccccccccccc}
\hline
\hline
NOAA & \multicolumn{3}{c}{Maximum} & & \multicolumn{3}{c}{Mininum} & & \multicolumn{3}{c}{Longitudes where $CC>0.6$}\\
\cline{2-4} \cline{6-8}\cline{10-12}
No. & $V_x$ & $V_y$ & $V$ &  & $V_x$ & $V_y$ & $V$ &  & $V_x$ & $V_y$ & $V$  \\
\hline
11158 & 0.99 & 0.99 & 0.99 &  & 0.43 & 0.59 & 0.41 &  &  - W54 &  - W58 & - W54 \\
11072 & 0.99 & 0.97 & 0.97 &  & 0.58 & 0.54 & 0.51 &  &  - W58 &  - W59 & - W58 \\
11084 & 0.94 & 0.90 & 0.93 &  & 0.24 & 0.23 & 0.38 &  & E50 - W41 & E57 - W45 & E50 - W46 \\
\hline
\end{tabular}
}
\end{table}

Table 2 lists some statistics on these correlation coefficients. We see that for the two emerging active regions (NOAA 11158 and NOAA 11072) the range where $CC>0.6$ extends to 54 degrees in the west. Even for the simple active region NOAA 11084 the regions where $CC>0.6$ cover from E50 to W41. All these indicate that the effect of non-radial magnetic fields on obtaining tangential velocities is relatively small, as long as the active region is not too far away from the disk center, for example, 40 degrees within the disk center.

In addition, a note to put here is that, on obtaining the tangential velocity by DAVE code we have used the constant area map projections as mentioned in Section 2. Welsch et. al (2009) pointed out that such projections do not preserve direction, in which case the velocities $V_x$ and $V_y$ probably contain distortions.  Welsch et. al (2009) argued that one should use a conformal map, such as the Mercator projection, to ensure that the velocity directions are determined correctly. On the other hand, Liu and Schuck (2012) argue that if the active region being analyzed represents a small part of the surface, these distortions are probably not important. Since we are analyzing active regions as Liu and Schuck (2012) did, we expect the distortions on the obtained velocity maps to be small.

\subsection{Effect on measuring the magnetic helicity transport}

\subsubsection{On the magnetic helicity injection rate}

Using Equation (4) we can calculate the helicity injection rate. Again, we are interested in comparing the two types of helicity injection rates, one obtained using $B_r$ magnetograms, denoted as $dh/dt$ hereafter, and one obtained using estimated $B_r^{\prime}$ magnetograms, denoted as $dh^{\prime}/dt$ hereafter. As an example, Figure 9 gives the distributions of obtained magnetic helicity injection rates for NOAA 11158. Still the top panels show the distributions when the active region is near the disk center and the bottom panels when the active region is $45^o$ from the disk center.

From the top two panels of Figure 9 we see that the two distributions look very similar. The total magnetic helicity injection rate for the top-left panel is $dh^{\prime}/dt = 6.59\times10^{37}Mx^{2}s^{-1}$. For the top-right panel $dh/dt$ is $7.10\times10^{37}Mx^{2}s^{-1}$. The difference is $(7.10-6.59)/7.10=7\%$. However, when the active region is $45^o$ from the disk center, the difference between $dh^{\prime}/dt$ and $dh/dt$ become evident, as shown in the bottom two panels of Figure 9.
The two red squares show where they differ most. We see in this subregion even the signs of helicity injection rate are different. This reminds us that we need to be careful in interpreting the distribution of helicity injection rate map, particularly if we are using estimated $B_r^{\prime}$ magnetograms. The total helicity injection rate in the bottom-left panel is $dh^{\prime}/dt = 3.90\times10^{37}Mx^{2}s^{-1}$ and $dh/dt = 2.75\times10^{37}Mx^{2}s^{-1}$ for the bottom-right panel. The difference between these two is now $(3.90-2.75)/2.75=42\%$.

This phenomenon exists for all three active regions, as can be seen from the middle panels in Figure 10 (for NOAA 11158), Figure 11 (NOAA 11072) and Figure 12 (NOAA 11084). We see there that when the active regions are near the disk center, the values of $dh^{\prime}/dt$ (presented by dotted lines) and $dh/dt$ (presented by solid lines) are close to each other. But when the active regions are far away from the disk center, even the signs of $dh^{\prime}/dt$ and $dh/dt$ could be different.

Similarly as in Figure 8, to show the correlation between $dh/dt$ and $dh^{\prime}/dt$ more quantitatively, we have calculated the correlation coefficients between the two helicity injection rates for the three active regions during the entire passage across the disk. The results are presented in Figure 13. We see that, unlike the correlation coefficients between the tangential velocities, the correlation coefficients between the two helicity injection rates decrease very quickly as the active regions move away from the disk center. The value of the correlation coefficient can even become negative when the active region is around $60^o$ from the disk center.

\bigskip
\begin{table}[htbp]
\centering
\centerline{\footnotesize \textbf{Table 3.} Statistics on the correlation coefficients between $dh/dt$ and $dh^{\prime}/dt$}
\label{tab1}
{\scriptsize
\begin{tabular}{cccccc}
\hline
\hline
NOAA No. & Maximum & & Mininum & & Longitudes where $CC>0.6$\\
\hline
11158 & 0.98 &  & 0.10 &  & - W41 \\
11072 & 0.96 &  & -0.01 &  & - W34 \\
11084 & 0.85 &  & -0.10 &  & E41 - W24\\
\hline
\end{tabular}
}
\end{table}

Similar to Table 2, we list some statistics on the correlation coefficients between $dh/dt$ and $dh^{\prime}/dt$ in Table 3. We see here that for the helicity injection rates the longitudes where $CC>0.6$ shrink to within about 25 degrees from the disk center, which means that the effect of non-radial magnetic field on the calculation of magnetic helicity injection rate is relatively strong.

\subsubsection{On the amount of accumulated magnetic helicity}

Using Equation (5) we can calculate the amount of accumulated magnetic helicity, from the starting point of observation, for the three active regions. The results are presented in the bottom panels of Figures 10, 11 and 12 for NOAA 11158, NOAA 11072 and NOAA 11084 respectively.  Here $H$ denotes for the amount of accumulated magnetic helicity calculated using $B_r$ magnetograms and $H^{\prime}$ for the amount of accumulated magnetic helicity calculated using $B^{\prime}_r$ magnetograms.

We see from these panels that the differences between $H$ and $H^{\prime}$ are small, except for NOAA 11072 when locating 45 degrees further in the west. This shows that the effect of non-radial magnetic field on the measurement of accumulated magnetic helicity becomes less significant, as the amount of accumulated magnetic helicity comes mostly from the moments when the active regions are not far away from the disk center.

\bigskip
\begin{table}[htbp]
\centering
\centerline{\footnotesize \textbf{Table 4.} A comparison of the two types of helicity calculations near the western limb}
\label{tab1}
{\scriptsize
\begin{tabular}{cccccccccccccccccc}
\hline
\hline
NOAA  & \multicolumn{5}{c}{W40} & & \multicolumn{5}{c}{W50} & & \multicolumn{5}{c}{W60}\\
\cline{2-6}\cline{8-12}\cline{14-18}
No. &$\frac{dH}{dt}$&$\frac{dH^\prime}{dt}$&$H$&$H^\prime$&$\frac{|\triangle H|}{H}$&
&$\frac{dH}{dt}$&$\frac{dH^\prime}{dt}$&$H$&$H^\prime$&$\frac{|\triangle H|}{H}$&
&$\frac{dH}{dt}$&$\frac{dH^\prime}{dt}$&$H$&$H^\prime$&$\frac{|\triangle H|}{H}$\\
\hline
11158  &41.6&19.4&24.6&20.7&16\%&  &-23.4&4.44&25.4&21.5&15\%&  &-54.7&-8.06&22.4&21.1&6\%\\
11072  &0.29&-4.54&-1.80&-2.03&13\%&  &5.57&0.28&-1.64&-2.18&33\%&  &5.80&3.23&-1.38&-2.35&70\%\\
11084  &1.29&-0.43&2.46&2.48&1\%&  &-0.89&-1.32&2.50&2.50&0&  &1.11&-5.92&2.58&2.50&3\%\\
\hline
\end{tabular}
\begin{flushleft}
Note: The unit for $\frac{dH}{dt}$  and $\frac{dH^\prime}{dt}$ is $10^{36}Mx^{2}s^{-1}$, for $H$  and $H^\prime$ $10^{42}Mx^{2}$.
\end{flushleft}}
\end{table}

A more quantitative comparison is given in Table 4. We see that at the position of W40, even though the helicity injection rates of $dh/dt$ and $dh^{\prime}/dt$ are obviously different, the difference between $H$ and $H^{\prime}$ are not very large. The largest difference between $H$ and $H^{\prime}$ at W40 is $16\%$. Interesting is that for active regions NOAA 11158 and NOAA 11084 even at 60 degrees from the disk center, the differences between $H$ and $H^{\prime}$ are still smaller than $6\%$.

These indicate that if the observation is done not far away from the disk center, for example within 40 or 45 degrees from the disk center, the effect of non-radial magnetic field on the calculation of accumulated magnetic helicity is not severe.


\section{Conclusion and discussion}

We have used HMI/SDO magnetic field data to investigate the effect of non-radial magnetic field on measuring magnetic helicity transport across solar photosphere. Three active regions are studied. They are NOAA 11072, NOAA 11084 and NOAA 11158.

First, we compared the differences between the true radial magnetic field $B_r$ and the estimated radial field obtained by $B^{\prime}_r= B_l/cos\theta$ based on the assumption that magnetic field is radial. The comparison shows that the radial assumption is not valid and can bring in significant distortion on the magnetic structure when the active region is far away from the disk center.

Then we studied the effect of the non-radial magnetic field on estimating tangential velocity using the method of DAVE. Interestingly we found that the effect is relative small. The main reason might be that, though the structure of the estimated radial magnetic field obtained based on the radial assumption is obviously different from that of the true radial field when the active region is away from disk center, the information on the field-line footpoints movement is still preserved.

Finally, we discussed the effect of non-radial magnetic field on estimating the magnetic helicity transport. When the active region is far away from the disk center, the radial assumption can bring in much distortion on the distribution of helicity injection rate and even the sign of the injection rate can be opposite to that of the true one. However, if we only consider the amount of accumulated magnetic helicity, the effect of non-radial magnetic field then becomes minor, as long as our calculation is done within 40 or 45 degrees from the disk center.

\acknowledgements

We thank Yang Liu (Stanford University) for helps in providing DAVE4VM package which includes DAVE programs and in treating HMI/SDO data. We also thank the referee for helpful comments and suggestions that improved the presentation of the paper. This work was partially supported by the National Natural Science Foundation of China (Grants No. 11125314 and No. 11221063), the National Basic Research Program of MOST (Grant No. 2011CB811401), the Knowledge Innovation Program (Grant No. KJCX2-EW-T07) and the Strategic Priority Research Program (Grant No. XDB09000000) of the Chinese Academy of Sciences (CAS), the Young Researcher Grant of National Astronomical Observatories of CAS.


\newpage

\begin{figure}[!ht]
\centerline{\includegraphics[width=0.6\textwidth]{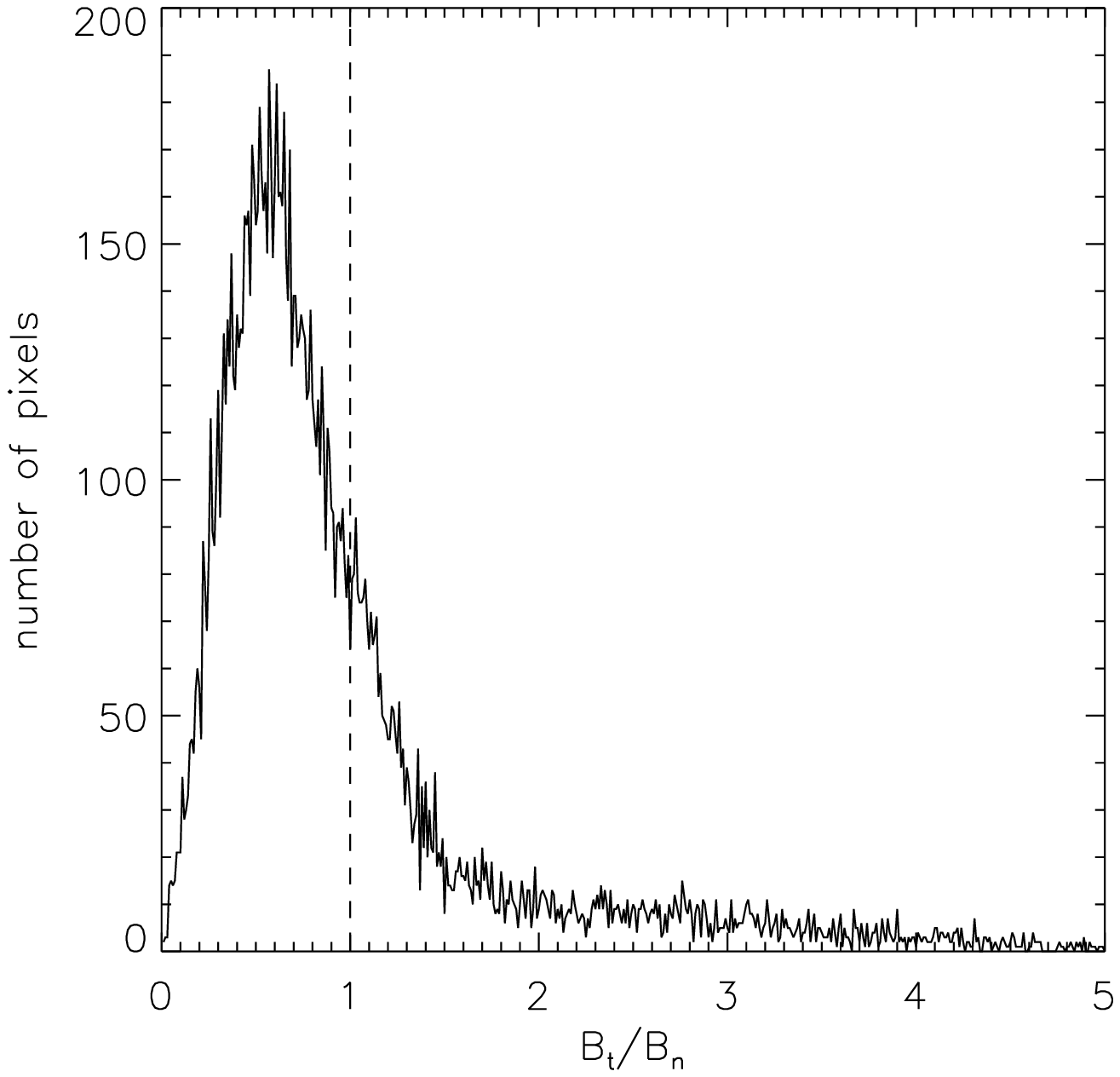}}
\caption{Histogram of the ratio between the strengths of tangential field and normal field in one magnetogram of AR11072, obtained at 16:36 UT on 2010 May 25. The horizontal axis is the value of the ratio, the vertical axis is the number of pixels.}
\end{figure}

\begin{figure}[!ht]
\centerline{\includegraphics[width=0.95\textwidth]{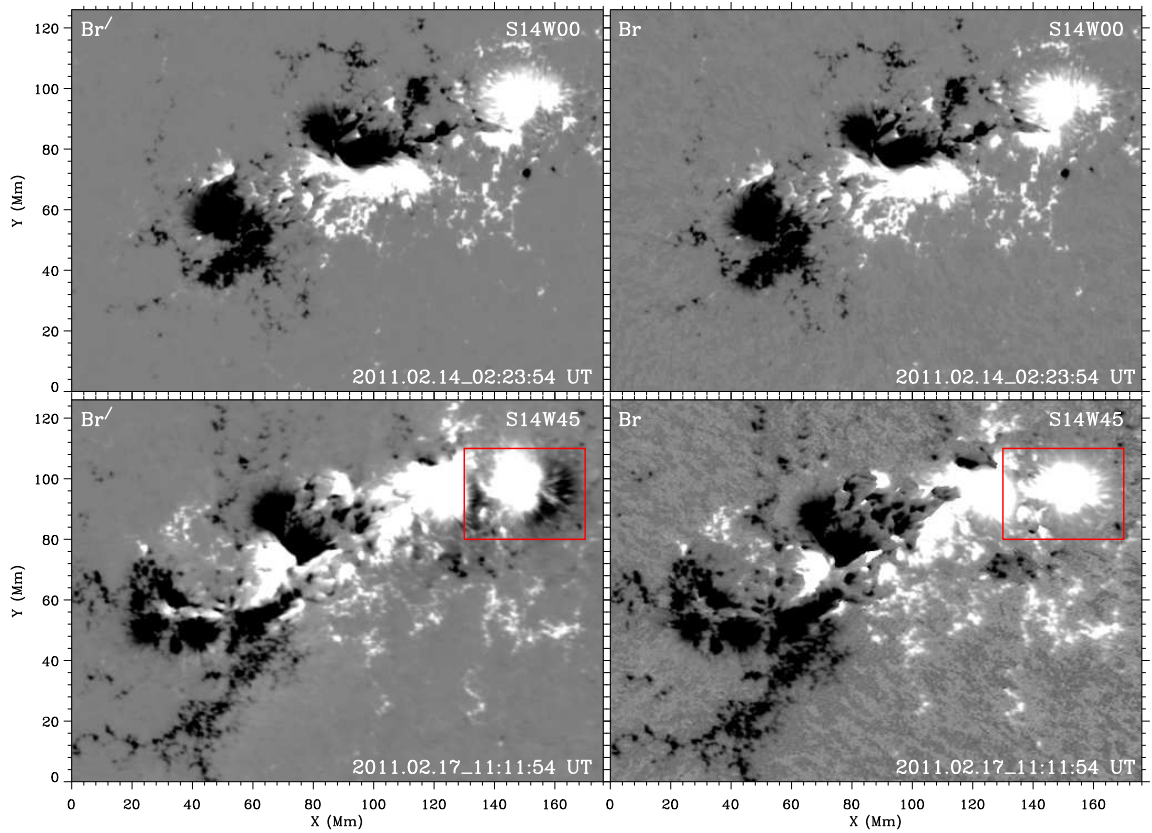}}
\caption{Maps of the radial magnetic field ($B_r$, right panels) and the derived radial magnetic field ($B_r^{\prime}$, left panels) of NOAA 11158. Top panels show the fields when the active region is near the disk center; bottom panels when the active region is $45^o$ from the disk center. The red square outlines the region where $B_r$ and $B_r^{\prime}$ differ most.}
\end{figure}

\begin{figure}[!ht]
\centerline{\includegraphics[width=0.95\textwidth]{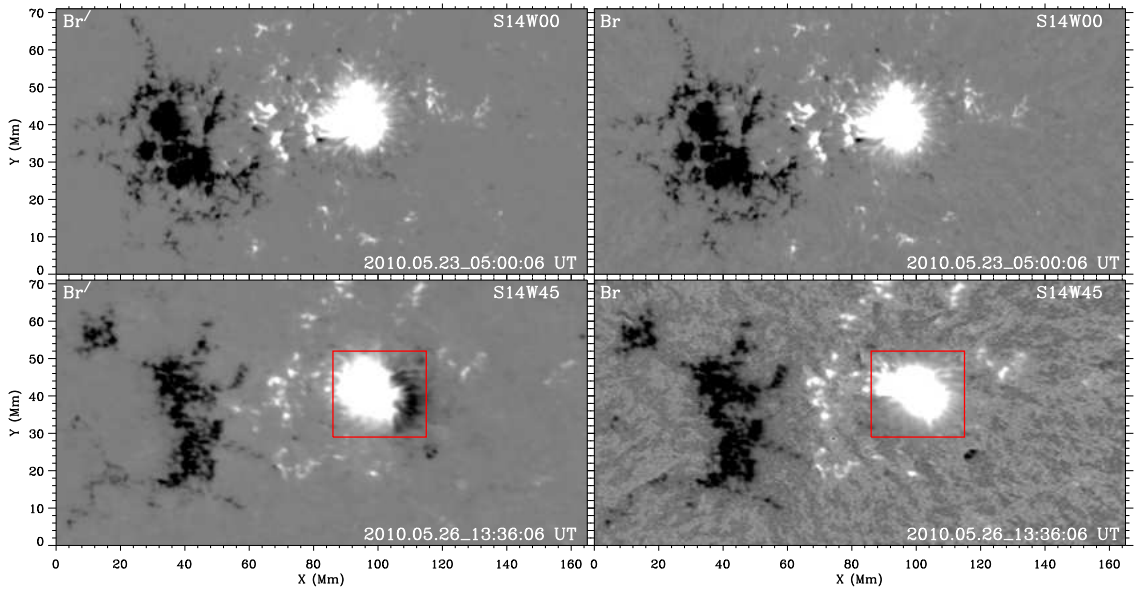}}
\caption{Similar as Figure 2, maps of the radial magnetic field ($B_r$) and the derived radial magnetic field ($B_r^{\prime}$) of NOAA 11072 in two positions. The red square outlines the region where $B_r$ and $B_r^{\prime}$ differ most.}
\end{figure}

\begin{figure}[!ht]
\centerline{\includegraphics[width=0.7\textwidth]{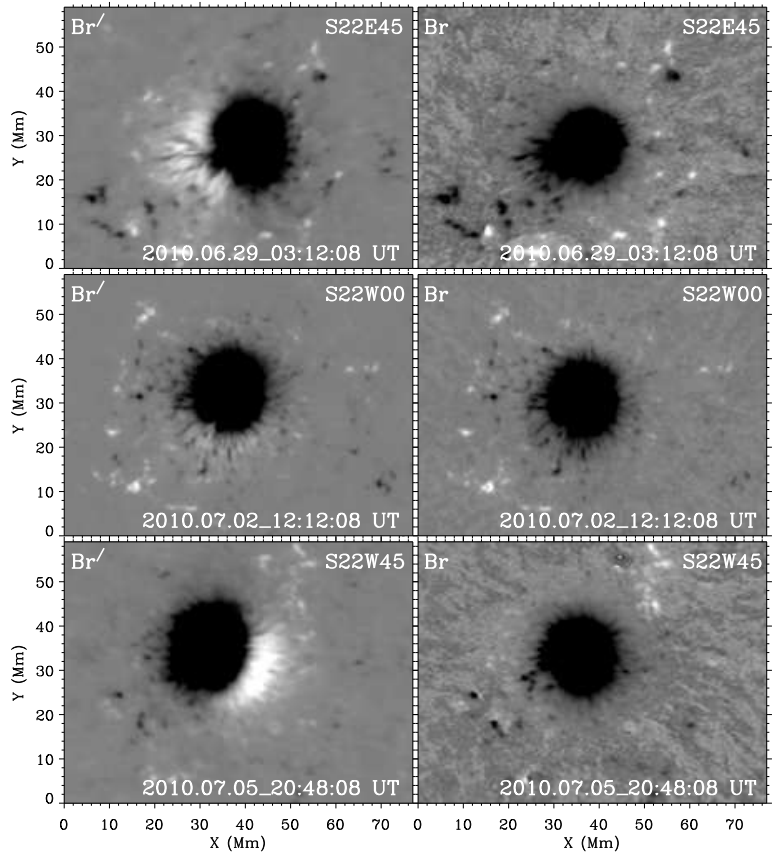}}
\caption{Same as Figure 2, but for NOAA 11084. Here middle panels show the fields when the active region is near the disk center; both top and bottom panels show the fields when the active region is $45^o$ from the disk center, top panels in the east and bottom panels in the west.}
\end{figure}

\begin{figure}[!ht]
\centerline{\includegraphics[width=0.9\textwidth]{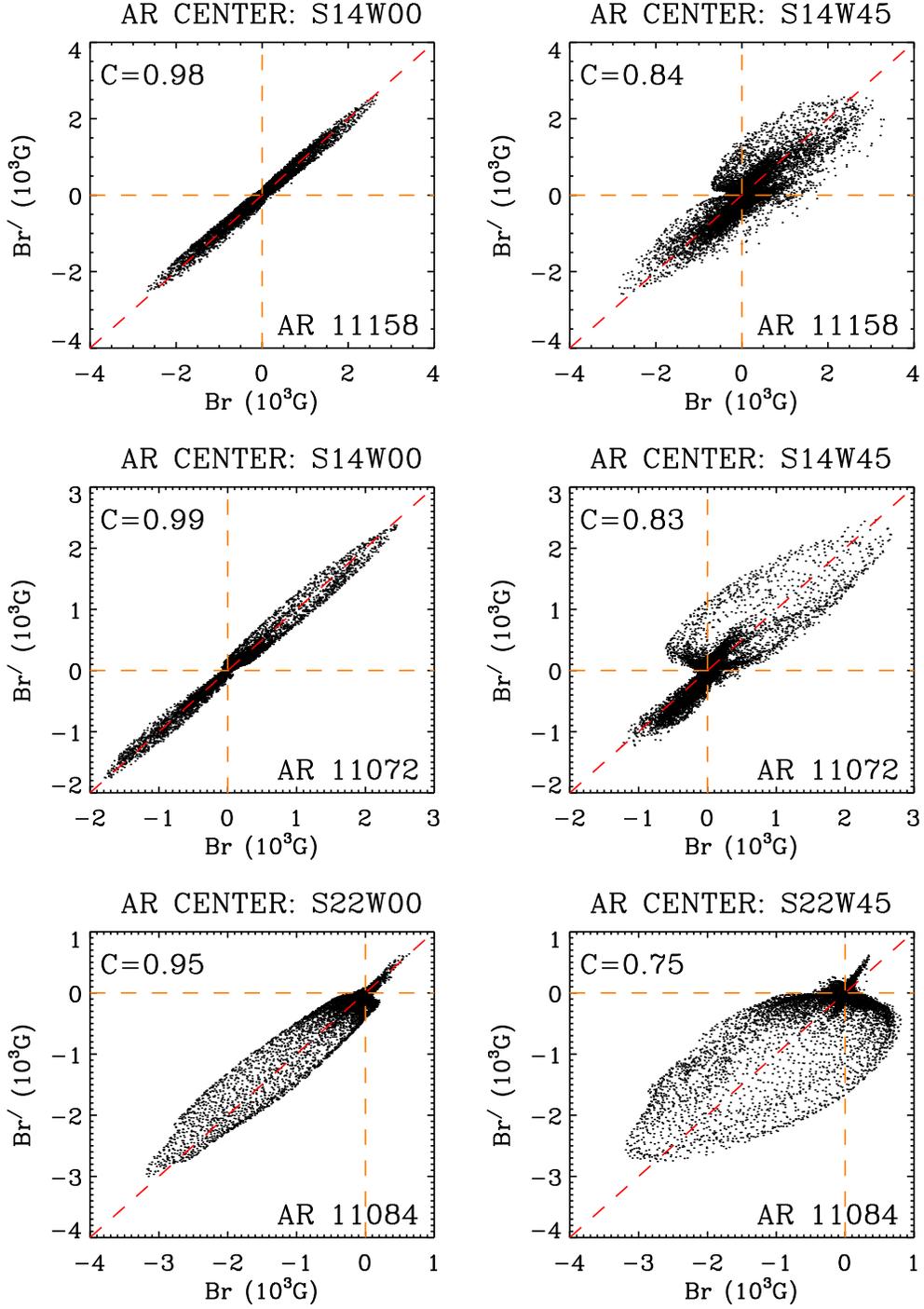}}
\caption{\small{Scatter plots of $B_r$ vs. $B_r^{\prime}$ for NOAA 11158 (top panels), NOAA 11072 (middle panels) and NOAA 11084 (bottom panels). Left panels for active regions near the disk center and right panels for regions close to the limb.}}
\end{figure}

\begin{figure}[!ht]
\centerline{\includegraphics[width=0.95\textwidth]{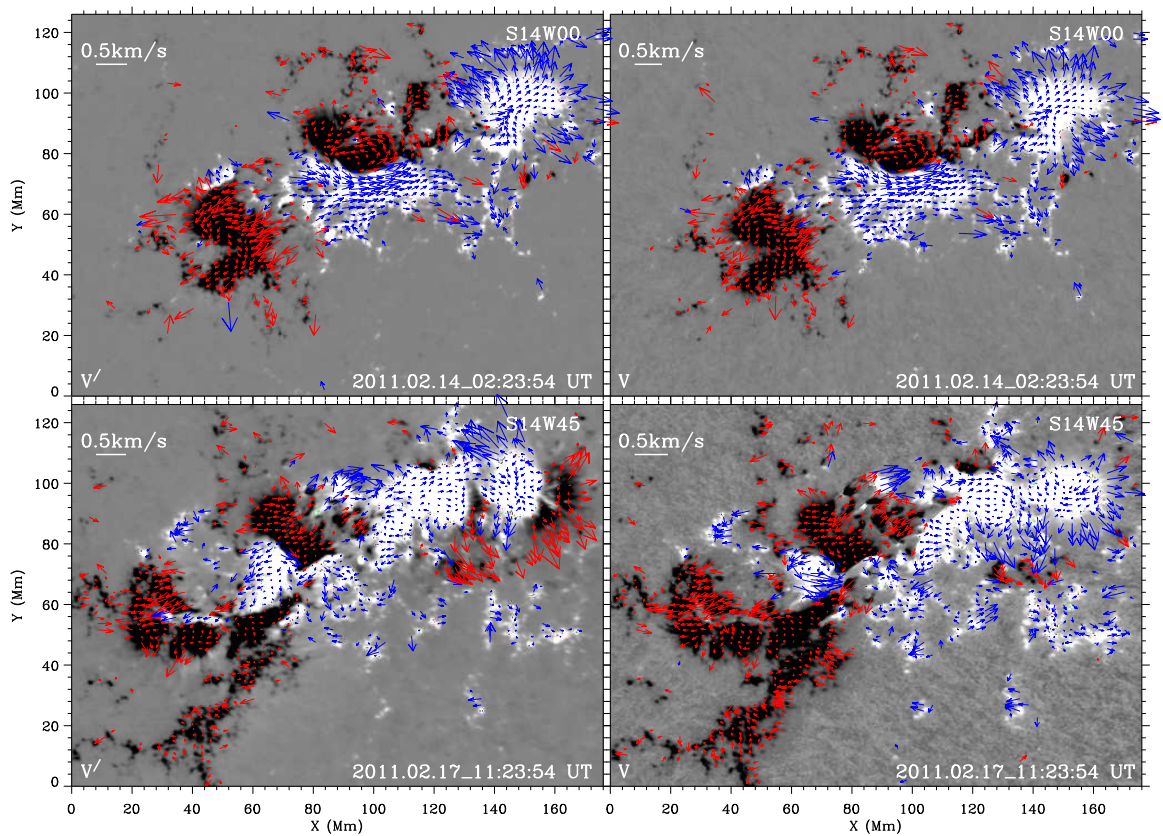}}
\caption{Tangential velocity field (arrows) plotted over the radial magnetic field ($B_r$ or $B_r^{\prime}$, black-white image) of NOAA 11158. Blue arrows for regions where $B_r>0$ or $B_r^{\prime}>0$, red for $B_r<0$ or $B_r^{\prime}< 0$. Right panels show velocities obtained by using $B_r$ magnetograms; Left panels are velocities obtained by using $B_r^{\prime}$ magnetograms. Similar to Figure 2, top panels show the fields when the active region is near the disk center and bottom panels the fields when the active region is $45^o$ from the disk center.}
\end{figure}

\begin{figure}[!ht]
\centerline{\includegraphics[width=0.85\textwidth]{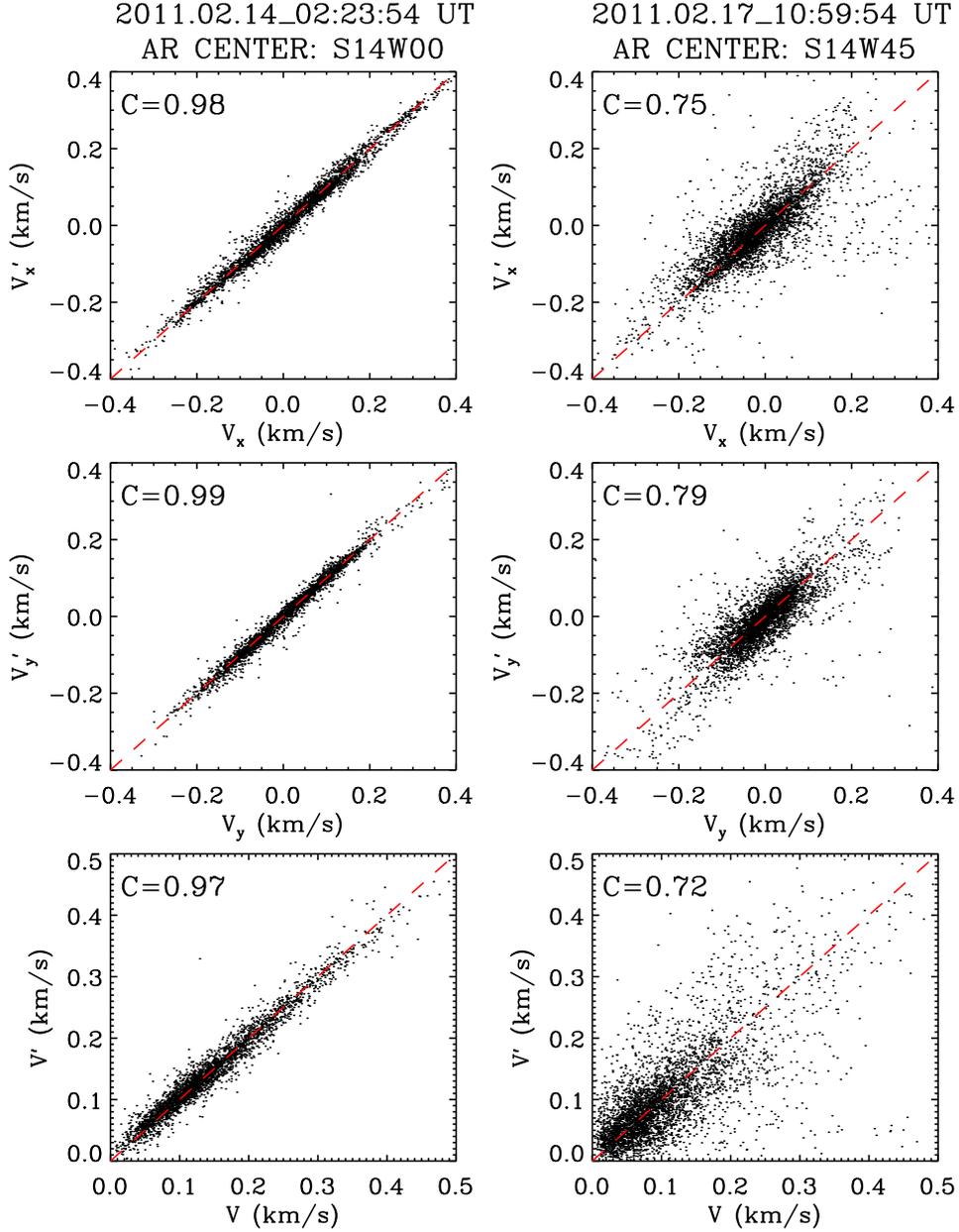}}
\caption{\small{The correlation between the two types of velocity maps shown in Figure 5, one obtained using $B_r^{\prime}$ and the other using $B_r$ magnetograms. Left panels for the moment when the active region is near the disk center, right panels when the active region is $45^o$ in the west. Top panels for the $V_x$ (positive in the direction of solar rotation), middle panels for $V_y$ (positive from south to north), and the bottom panels for $V$ (the magnitude of tangential velocity). In the top-left corner of each panel shows the correlation coefficient between the two quantities plotted in each panel.}}
\end{figure}

\begin{figure}[!ht]
\centerline{\includegraphics[width=0.90\textwidth]{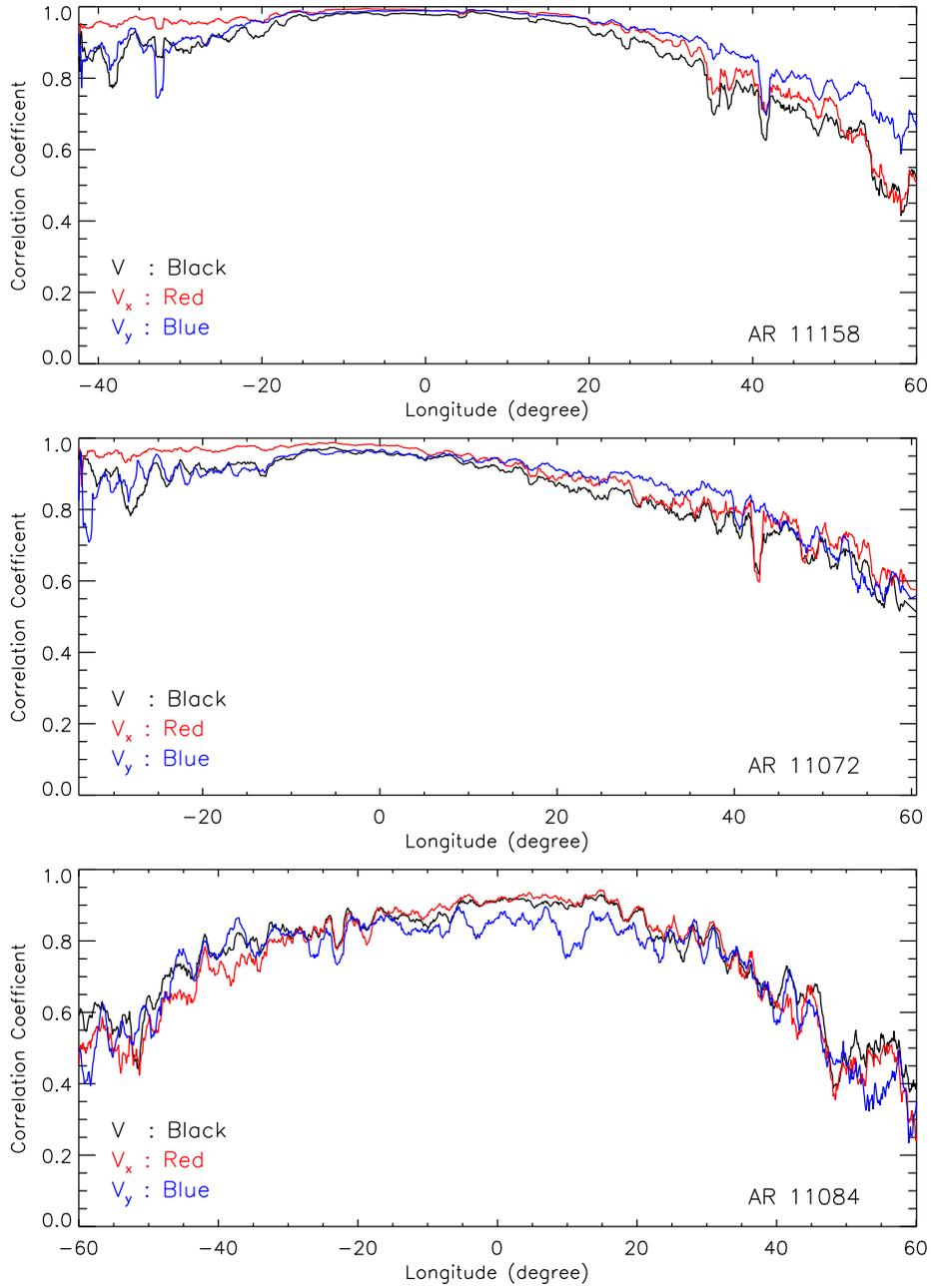}}
\caption{\small{Correlation coefficients between velocities obtained using $B_r$ magnetograms and those obtained using $B_r^{\prime}$ magnetograms, during the entire passage across the disk, for NOAA 11158 (top panel), NOAA 11172 (middle panel) and NOAA 11084 (bottom panel). Red lines for $V_x$, blue for $V_y$ and black for $V$.}}
\end{figure}

\begin{figure}[!ht]
\centerline{\includegraphics[width=0.95\textwidth]{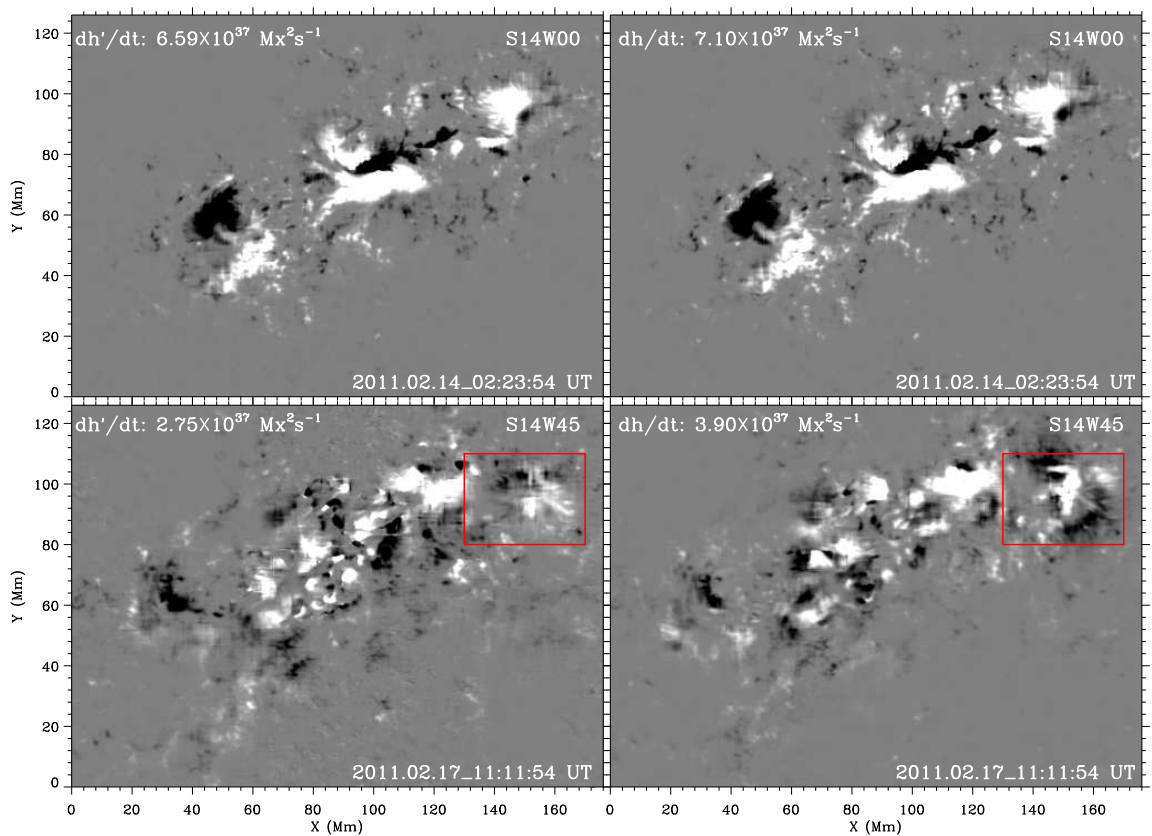}}
\caption{The distribution of magnetic helicity injection rate for NOAA 11158. Right panels are $dh/dt$ calculated by using $B_r$ magnetograms, left panels are $dh^{\prime}/dt$ by using $B_r^{\prime}$ magnetograms. Again, top panels for the field near the disk center and bottom panels for the field $45^o$ from the disk center. Areas marked by the red box are where $dh/dt$ and $dh^{\prime}/dt$ differ most.}
\end{figure}

\begin{figure}[!ht]
\centerline{\includegraphics[width=0.9\textwidth]{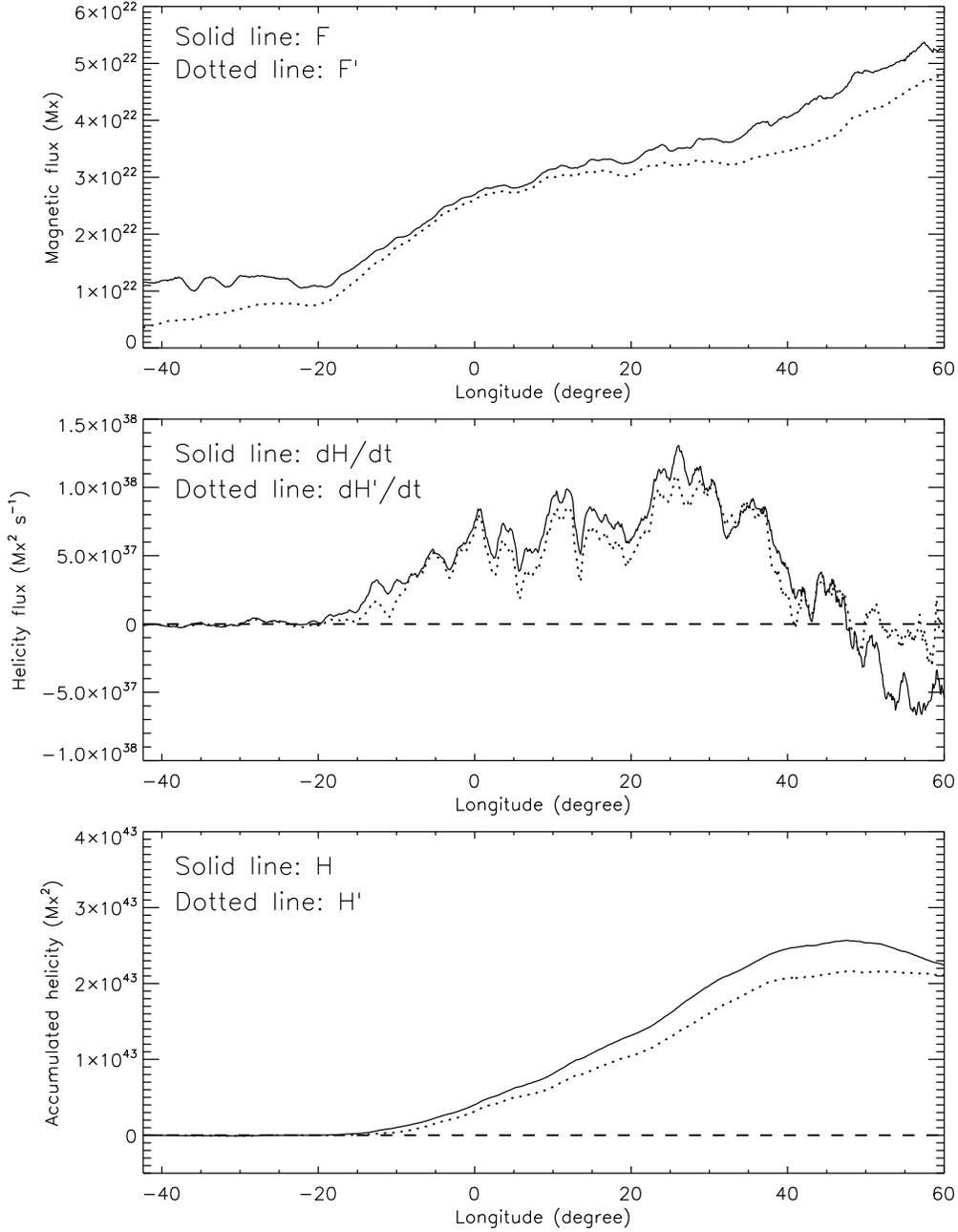}}
\caption{Temporal profiles of unsigned magnetic flux (top panel), magnetic helicity injection rate (middle panel) and accumulated magnetic helicity (bottom panel) of NOAA 11158. Solid lines for quantities obtained using $B_r$ magnetograms, dotted lines for using $B^{\prime}_r$.}
\end{figure}

\begin{figure}[!ht]
\centerline{\includegraphics[width=0.9\textwidth]{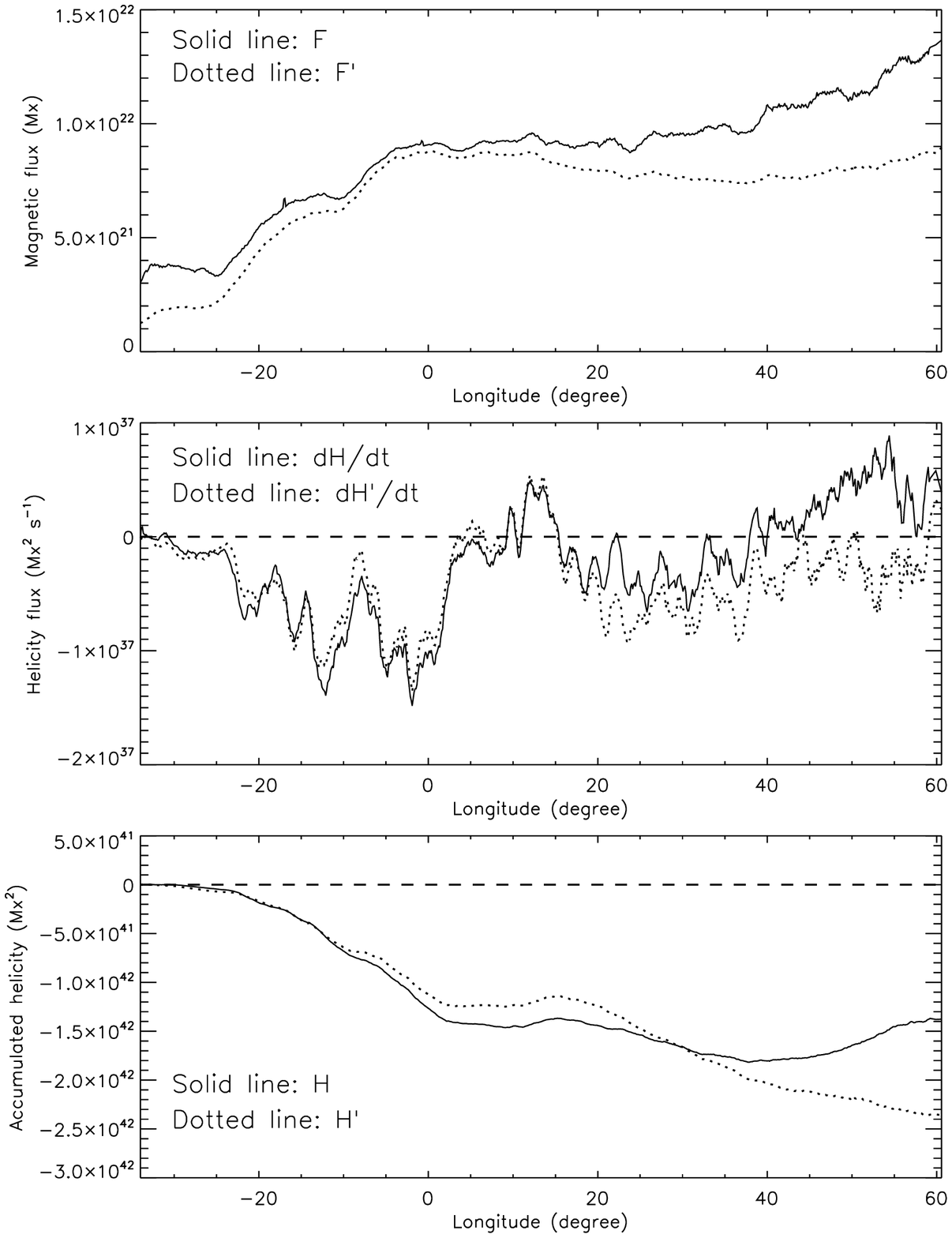}}
\caption{Same as Fig.10, temporal profiles of unsigned magnetic flux (top panel), magnetic helicity injection rate (middle panel) and accumulated magnetic helicity (bottom panel) for NOAA 11072.}
\end{figure}

\begin{figure}[!ht]
\centerline{\includegraphics[width=0.9\textwidth]{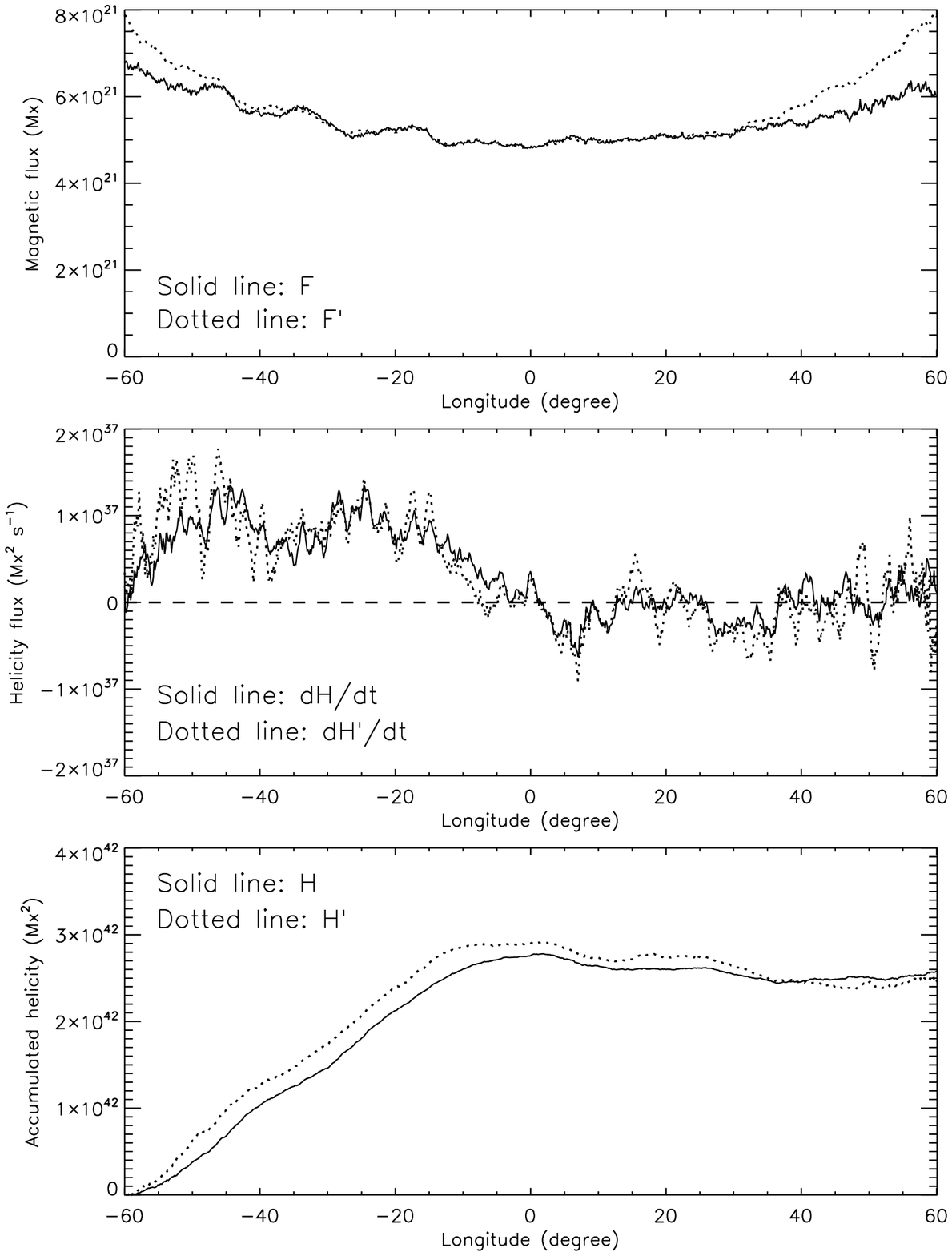}}
\caption{Same as Fig.10, temporal profiles of unsigned magnetic flux (top panel), magnetic helicity injection rate (middle panel) and accumulated magnetic helicity (bottom panel) for NOAA 11084.}
\end{figure}

\begin{figure}[!ht]
\centerline{\includegraphics[width=0.95\textwidth]{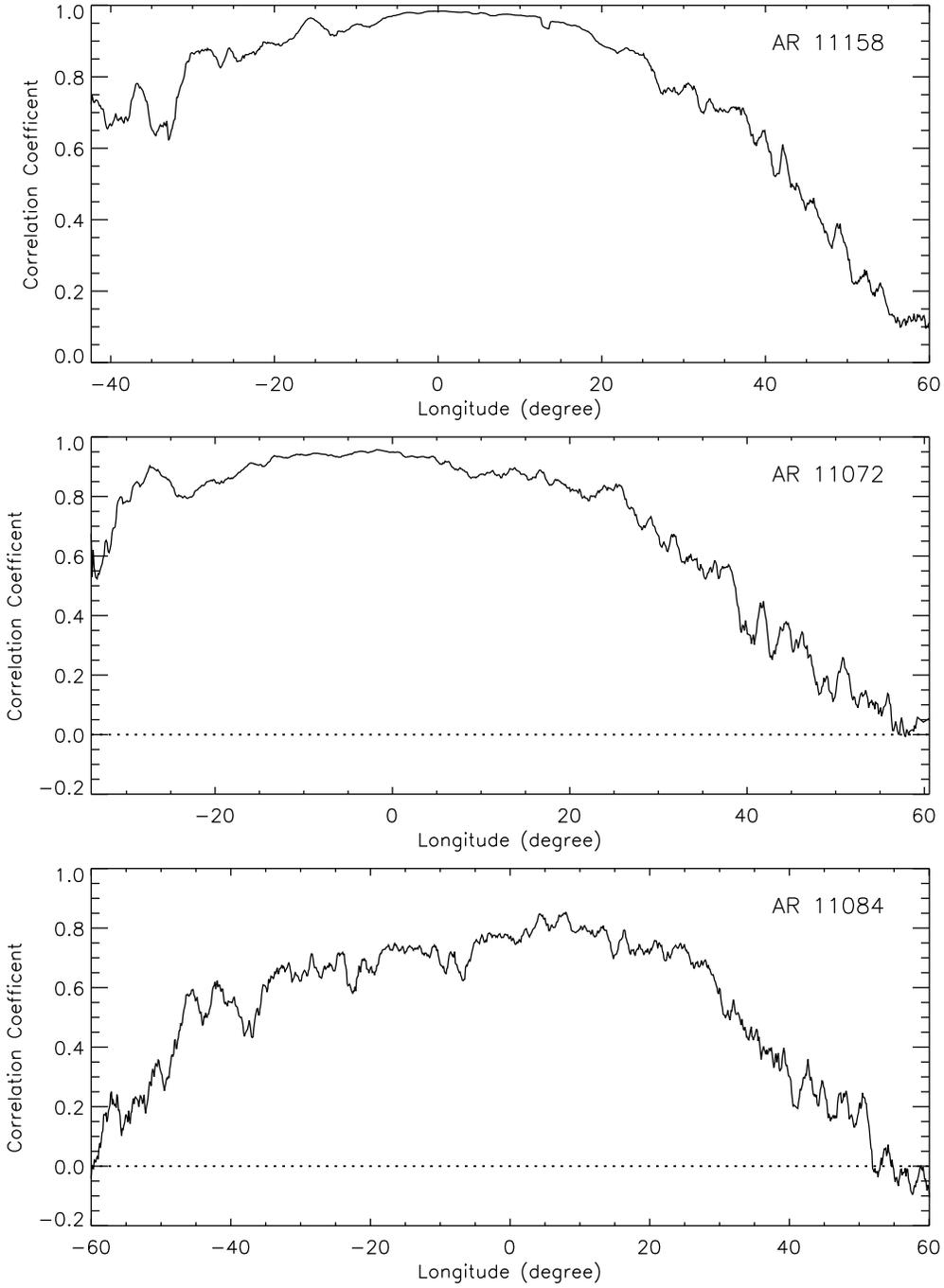}}
\caption{Similar to Fig.8 but for correlation coefficients between the helicity injection rates during the entire passage cross the disk for the three active regions.}
\end{figure}

\end{document}